\documentclass[10pt,conference]{IEEEtran} 
\IEEEoverridecommandlockouts
\usepackage{acronym}
\usepackage{hyperref}
\usepackage{makecell}
\usepackage{pifont}
\usepackage{cite}
\usepackage{amsmath,amssymb,amsfonts}
\usepackage{algorithmic}
\usepackage{graphicx}
\usepackage{textcomp}
\usepackage{xcolor}
\usepackage{tikz}

\DeclareRobustCommand*{\IEEEauthorrefmark}[1]{%
  \raisebox{0pt}[0pt][0pt]{\textsuperscript{\footnotesize\ensuremath{#1}}}}
  
\def\BibTeX{{\rm B\kern-.05em{\sc i\kern-.025em b}\kern-.08em
    T\kern-.1667em\lower.7ex\hbox{E}\kern-.125emX}}

\begin{document}
\newcommand\copyrighttext{%
  \footnotesize \textcopyright 2026 IEEE. Personal use of this material is permitted.
  Permission from IEEE must be obtained for all other uses, in any current or future
  media, including reprinting/republishing this material for advertising or promotional
  purposes, creating new collective works, for resale or redistribution to servers or
  lists, or reuse of any copyrighted component of this work in other works.}
  
\newcommand\copyrightnotice{%
\begin{tikzpicture}[remember picture,overlay]
\node[anchor=south,yshift=10pt] at (current page.south) {\fbox{\parbox{\dimexpr\textwidth-\fboxsep-\fboxrule\relax}{\copyrighttext}}};
\end{tikzpicture}%
}
\acrodef{QPU}[QPU]{Quantum Processing Unit}
\acrodef{QEC}[QEC]{Quantum Error Correction}
\acrodef{ASIC}[ASIC]{Application-Specific Integrated Circuit}
\acrodef{FPGA}[FPGA]{Field-Programmable Gate Array}
\acrodef{MF}[MF]{Matched Filter}
\acrodef{RMF}[RMF]{Relaxation Matched Filter}
\acrodef{EMF}[EMF]{Excitation Matched Filter}
\acrodef{SVM}[SVM]{Support Vector Machine}
\acrodef{ML}[ML]{Machine Learning}
\acrodef{NN}[NN]{Neural Network}
\acrodef{FNN}[FNN]{Feedforward Neural Network}
\acrodef{SSM}[SSM]{Selective State-Space Model}
\acrodef{FP}[FP]{False Positive}
\acrodef{FN}[FN]{False Negative}
\acrodef{TP}[TP]{True Positive}
\acrodef{ADC}[ADC]{Analog-to-Digital Converter}
\acrodef{DAC}[DAC]{Digital-to-Analog Converter}
\acrodef{I}[I]{In-Phase}
\acrodef{Q}[Q]{Quadrature}
\acrodef{MCM}[MCM]{Mid-Circuit Measurement}
\acrodef{MTV}[MTV]{Mean Trace Value}
\acrodef{KNN}[KNN]{k-Nearest Neighbors}
\acrodef{FTQC}[FTQC]{Fault-Tolerant Quantum Computing}

\title{Multi-Stage Mamba-Based Architecture for Fast and Scalable Superconducting Qubit Readout
\thanks{This work was funded by the German Federal Ministry of Education and
Research (BMBF) under the funding program Quantum Technologies - From
Basic Research to Market under contract number 13N16087, as well as from
the Munich Quantum Valley (MQV), which is supported by the Bavarian State
Government with funds from the Hightech Agenda Bayern.}
}

\author{\IEEEauthorblockN{Luca Otting\IEEEauthorrefmark{1}, Xiaorang Guo\IEEEauthorrefmark{1}, Emmanouil Giortamis\IEEEauthorrefmark{2}, Benjamin Lienhard\IEEEauthorrefmark{3,4},\\ Pramod Bhatotia\IEEEauthorrefmark{2} and Martin Schulz\IEEEauthorrefmark{1}}

\IEEEauthorblockA{\IEEEauthorrefmark{1}Chair of Computer Architecture and Parallel Systems, Technical University of Munich, Garching, Germany \\ 
\IEEEauthorrefmark{2}Systems Research Group, Technical University of Munich, Garching, Germany \\ 
\IEEEauthorrefmark{3}TUM School of Natural Sciences, Technical University of Munich, Garching, Germany \\
\IEEEauthorrefmark{4}Walther-Meißner-Institut, Garching, Germany\\
Email: \{luca.otting, xiaorang.guo, emmanouil.giortamis, benjamin.lienhard, pramod.bhatotia, martin.w.j.schulz\}@tum.de}}

\maketitle
\copyrightnotice
\begin{abstract}
Reliable qubit readout is a critical bottleneck toward fault-tolerant quantum computing (FTQC). In superconducting quantum processors, readout operations are both error-prone and high-latency. These challenges become more severe in frequency-multiplexed architectures, where signal crosstalk among neighboring qubits significantly degrades readout fidelity. Existing machine learning (ML)-based approaches rely on feed-forward neural networks (FNNs) that suffer from large parameter sizes and lack an end-to-end network that jointly addresses relaxation errors and discriminates qubit states.

In this work, we present a multi-stage qubit state discriminator based on the Mamba model, which enables efficient sequence modeling with linear complexity. The first stage performs initial state discrimination, followed by a refinement stage that identifies and mitigates relaxation-induced errors. Our lightweight model achieves a geometric mean readout fidelity of 0.906, outperforming the best-reported state-of-the-art method while reducing parameter size by 49.6\%; our optimal model further reaches 0.911. Both models remain robust across varying input trace lengths, maintaining a high fidelity of 0.893 at readout durations as short as 500 $ns$, achieving up to a 26\% reduction in logical error rate over prior work in quantum error correction (QEC).

\end{abstract}


\section{Introduction}
\label{sec:intro}
Quantum computing has recently shown great potential for solving classically intractable tasks, such as chemistry simulations~\cite{WEIDMAN2024102105}, optimization~\cite{Jiang_op23}, and cryptography problems~\cite{Amr25}. Among all the physical modalities that are used to build the \acp{QPU}, the superconducting qubit is regarded as one of the most promising technologies due to its fast gate operation and compatibility with existing semiconductor fabrication techniques~\cite{abughanem2025superconducting}. Moreover, superconducting qubits have been extensively studied in the context of \ac{QEC}, including the construction of surface codes~\cite{zhao2022realization} and the development of efficient decoders~\cite{Vittal23}.

Qubit readout, also known as qubit state discrimination, is a fundamental operation in quantum computing, where quantum information encoded in a superposition state is converted into classical information represented by binary values '0' and '1'. However, the readout process in superconducting quantum computers remains the most error-prone and latency-critical stage~\cite{mude2025efficient,Benjamin2022}. Concretely, on quantum hardware such as the IBM Heron \ac{QPU}s \cite{ibm-devices}, measurement error rates range from $4\times10^{-3}$ to $3\times10^{-2}$, up to $12\times$ higher than 2-qubit gate errors. Furthermore, measurement durations of 1.5–2.5 $\mu s$ represent a $23-38\times$ latency increase relative to standard 2-qubit operations (68–88 ns), significantly impacting the feasibility of real-time feedback and overall circuit depth.

This limitation hinders the realization of practical quantum systems that rely on \ac{MCM} and \ac{QEC}, where accurate and fast feedback in the nanosecond range is typically required~\cite{rudinger2022characterizing}. 
In \ac{QEC} codes such as the surface code, the states of logical qubits are maintained through continuous rounds of syndrome extraction. During the \ac{MCM} of these syndrome qubits, the data qubits sit idle. The longer a data qubit sits idle, the higher the chance it decoheres \cite{burnett2019decoherence}. If the readout takes too long, these accumulated qubit errors can have a significant impact on the overall error rate. \textit{We therefore require a readout architecture that minimizes latency while maximizing discrimination fidelity.}

In addition, considering the scalability of quantum computers, resource-efficient superconducting chips are designed to couple multiple qubit readout resonators to a single feedline via frequency multiplexing~\cite{chen2012multiplexed,Bakr_2025}. Yet, the readout signals in frequency-multiplexed qubit systems often suffer from crosstalk, making qubit state discrimination more challenging. \textit{Consequently, the development of precise and scalable readout algorithms for these architectures is essential.}



To date, \ac{ML}-based approaches are actively investigated as alternatives to classical discriminators, such as \acp{MF} \cite{MF} and \ac{SVM}~\cite{SVM}, due to their relatively high performance in frequency-multiplexed readout. Early studies utilize \acp{NN}~\cite{Benjamin2022,gautam2024low,Liu25QR} or combine them with variants of \acp{MF} (e.g., HERQULES~\cite{MF/RMF}). While these methods provide high readout fidelity, they rely on the simultaneous readout of all qubits and therefore \textit{cannot support \ac{MCM}, where individual (single) qubit readout is required.} To overcome this limitation, several recent works, including QubiCML~\cite{QubiCML}, and KLiNQ~\cite{Guo25,guo2026hardware} have been proposed. Although these approaches enable \ac{MCM}, they remain based on fully connected  \acp{FNN}, which inherently involve a large number of parameters, since the number of connections between two layers grows proportionally to the product of their neuron counts. As a result, they often sacrifice discrimination accuracy to reduce network complexity and improve scalability.

To address the drawback of \ac{FNN}, we adopt a \ac{SSM} called Mamba~\cite{gu2024mamba}, which has recently emerged as a promising architecture to deal with various downstream tasks, including sequence processing applications, and requires only linear computational complexity with the increase of the input sequence length~\cite{wei2025lightmamba}. Thus, \textit{the Mamba architecture is well-suited for analyzing qubit readout traces.}

In parallel, we analyze qubit readout traces to identify the relaxation errors, which are also studied in recent works~\cite{MF/RMF,mude2025efficient}.
Such relaxation events cause the qubit state to transition from '1' to '0' during measurement, leading to label inconsistency. In other words, traces that are labeled as '1' may in fact correspond to the ground state at the end of readout. 
Therefore, this issue should be addressed during readout, since the inconsistent labels can significantly degrade the discriminator’s performance and further cause error propagation, particularly in qubit calibration procedures.

Therefore, in this work, we propose a \textit{multi-stage architecture for qubit state discrimination}. In the first stage, we employ the Mamba model to perform the initial classification of the qubit state. In the second stage, we develop an error detector to post-correct the relaxation errors. Unlike prior works~\cite{mude2025efficient, MF/RMF}, which rely on specific \acp{MF}, we integrate this functionality directly into our neural network pipeline, enabling end-to-end correction without additional filtering overhead. 

Benefiting from the linear-time complexity of Mamba and our efficient architectural design, our most lightweight model achieves a substantial parameter reduction of 49.6\% and 99.4\% compared to the latest state-of-the-art method (MF-RMF-EMF)~\cite{mude2025efficient} and the baseline \ac{FNN}~\cite{Benjamin2022}, respectively. This lightweight design ensures scalability with respect to the number of qubits and enables further deployment on resource-limited hardware platforms. Moreover, it achieves high readout fidelities of 0.906 with a trace duration of 1~$\mu$s and 0.893 even with a trace as short as 500~$ns$, outperforming all related works to the best of our knowledge. Our optimal model can further push readout fidelity to 0.911 with a trace duration of 1~$\mu$s, offering a favorable accuracy–efficiency tradeoff for deployment scenarios with relaxed resource constraints. Additionally, we evaluate the impact of readout fidelity on \ac{QEC} performance. At a readout duration of 500~$ns$, our method achieves up to 26\% reduction in logical error rate compared to related works, highlighting its practical significance toward \ac{FTQC}.



Overall, the main contributions of this paper are summarized below:
\begin{itemize}
    \item We propose a multi-stage qubit-state discrimination framework that integrates qubit trace classification and relaxation-error correction into a unified pipeline.
    \item We adopt the Mamba architecture to process qubit readout traces, achieving significant parameter reduction and a high geometric-mean readout fidelity of 0.911.
    \item We reduce the trace length required for discrimination to 500 $ns$ while still maintaining a readout fidelity of 0.893.
    \item We propose a QEC-compatible readout model and achieve a logical error rate up to 26\% lower at a readout duration of 500 $ns$.
\end{itemize}

\section{Background}
In this section, we introduce the basic principles of qubit readout for superconducting qubits and the architecture of Mamba.
\begin{figure}
    \centering
    \includegraphics[width=\linewidth, trim=5mm 2mm 5mm 1mm, clip]{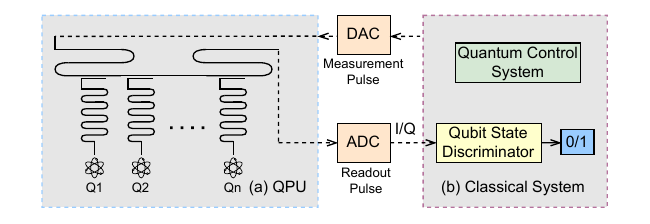}
    \caption{Readout pipeline of superconducting qubits. (a) On the \acp{QPU} side, the readout begins with a microwave pulse generated by the signal generator, then produces readout pulses that carry the qubit-state information. (b) On the classical side, the output signals are digitized into I/Q traces, which are then processed to determine the qubit states.}
    \label{fig:readout}
\end{figure}

\subsection{Superconducting Qubit Readout} 
Readout is the process of measuring a qubit's state, during which its superposition collapses into one of the two basis states: ground state ('0') or excited state ('1'). The entire readout pipeline spans multiple stages, from the analog domain to the digital domain, as illustrated in Figure~\ref{fig:readout}. Specifically, a readout pulse is first sent into the qubit’s resonator, producing an output signal that experiences a phase shift depending on the qubit’s state. This analog signal is then digitized by \acp{ADC}, and its \ac{I} and \ac{Q} components are extracted. Finally, the I/Q values are analyzed by classical processors or control hardware to discriminate between the ground and excited states~\cite{MF/RMF}. In this work, we focus on this final digital stage, known as single-shot qubit-state discrimination\cite{Benjamin2022}.

In addition, this work targets the multiplexed readout scenario, which is illustrated in Figure~\ref{fig:readout}(a)~ \cite{jerger2012frequency}. This architecture has become a key design choice for superconducting quantum processors due to constraints in chip area and the need for scalability. In this case, multiple qubit resonators are coupled to a shared microwave transmission line to reduce hardware overhead. However, the readout pulse synthesized for a specific resonator can also drive neighboring resonators, introducing crosstalk and consequently biasing the measured qubit states~\cite{Di_Giovanni_2025}. This interference brings an additional challenge to the qubit readout process, making accurate state discrimination more difficult.

\subsection{Mamba \ac{ML} Architectures}
Mamba~\cite{gu2024mamba} is a recently proposed \ac{SSM} in the \ac{ML} community that has demonstrated competitive performance with Transformers~\cite{Transformer} in sequence modeling and language tasks. Instead of having quadratic complexity in sequence length during training and requiring linearly growing cache memory during inference, Mamba achieves linear-time attention by maintaining a constant-sized hidden state, offering significant computational and memory efficiency. Structurally, Mamba extends the general discrete form of structured \acp{SSM}, where the hidden state $h_t$, input $x_t$, and output $y_t$ follow:
\begin{equation}
    h_t = A_t h_{t-1} + B_t x_t,
\end{equation}
\begin{equation}
    y_t = C_t^\mathrm{T} h_t.
\end{equation}
In contrast to linear time-invariant \acp{SSM}, Mamba allows the parameters $A_t$, $B_t$, and $C_t$ to vary with time, which gives Mamba the ability to focus on or ignore inputs at time step $t$~\cite{gu2024mamba}.

\begin{figure}
    \centering
    \includegraphics[width=0.95\linewidth, trim=7mm 5mm 7mm 7mm, clip]{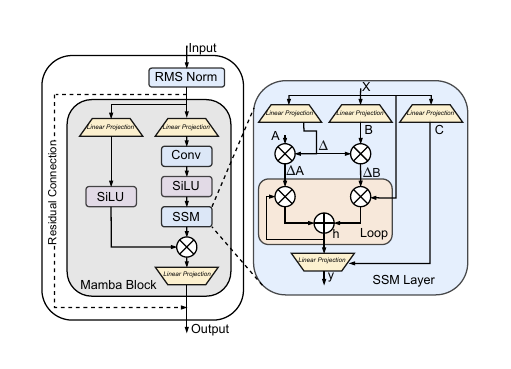}
    \caption{The basic structure of Mamba Model and its internal \ac{SSM} layer. The input is normalized and processed by a convolutional layer, followed by an SSM layer that performs sequence modeling through recurrent hidden-state (h) updates using learnable matrices A, B, and C. A residual connection merges the input and output of \ac{SSM} to form the final result.}
    \label{fig:mamba}
\end{figure}

The basic computational architecture of Mamba is shown in Figure~\ref{fig:mamba}. It processes inputs through linear and convolutional projections, applies an activation, and updates the state with the SSM~\cite{li2024marca}. Since each Mamba layer maintains an internal state that evolves over time, the model naturally captures temporal dependencies across readout time bins, making it particularly suitable for processing the quantum readout data.

\subsection{\ac{FTQC} via \ac{QEC}}
Realizing a practical quantum advantage hinges on fault-tolerant operation, as physical qubits are inherently susceptible to decoherence and gate errors that accumulate rapidly without active intervention \cite{fowler2012surface, gidney2021how, preskill1997faulttolerantquantumcomputation}. QEC addresses this by distributing a single logical qubit across an ensemble of physical qubits, encoding information in a subspace structured so that typical errors map the state into distinguishable, correctable error sectors \cite{Nielsen_Chuang_2010, Roffe_2019}. Crucially, identifying which sector the system has entered requires \ac{MCM} on ancilla qubits: these extract syndrome information without disturbing the encoded logical state, and must be repeated in successive cycles to track and suppress error propagation before it becomes irrecoverable \cite{decoding_2023, Roffe_2019}. Because the corrective action taken in each cycle is derived entirely from the syndrome bits produced by these measurements, the accuracy with which each ancilla readout is classified directly determines the reliability of the extracted syndrome, and therefore the overall logical error rate \cite{harper2025characterisingfailuremechanismserrorcorrected}. Improving qubit-state discrimination fidelity is thus not merely a hardware concern, but a prerequisite for effective \ac{QEC}.

\section{Mamba-Based Multi-Stage Architecture}
\label{sec:Architecture}
This section begins with the motivation and detailed design of the error detector and the main classification model, followed by the presentation of the proposed end-to-end multi-stage architecture.
\subsection{Integrated Error Detector}
\subsubsection{Challenge and Motivation}
Readout errors, such as relaxation or excitation events, can severely limit the performance of qubit state discrimination algorithms, as these events are particularly difficult to distinguish from normal traces. Since discrimination models are often designed to be lightweight for hardware deployment, their reduced parameter count can further lower accuracy, as the hard-to-classify error traces are often misclassified.

Current methods employ \ac{RMF}~\cite{MF/RMF,mude2025efficient} to help the model learn to better classify relaxation traces. However, the manual extraction of features using \acp{MF} poses the risk of losing valuable features by reducing readout traces to a single point in the complex plane. The loss of such information can lead to reduced readout accuracy in independent qubit readout scenarios~\cite{mude2025efficient}. Recognizing this issue, \textbf{our insight is that we need a reliable error detector that preserves full trace information to improve robustness, but can be seamlessly integrated into the \ac{NN} as well.}

\subsubsection{Approach}



\begin{figure}
    \centering
    \includegraphics[width=\linewidth]{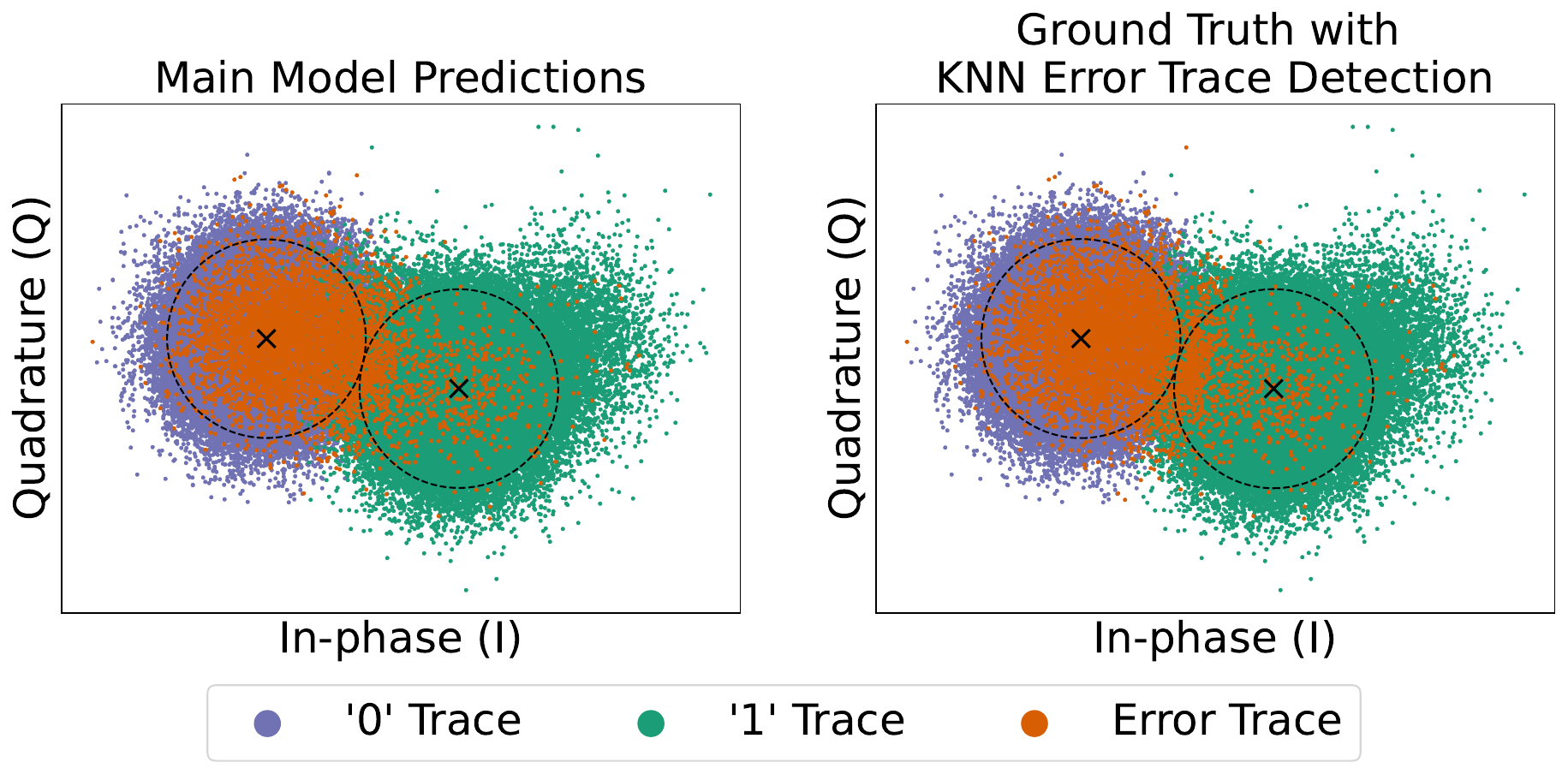}
    \caption{
    Comparison between the main model’s predicted trace clusters (left) and the pseudo-ground-truth labels obtained from the \ac{KNN}-based error detector (right).}
    \label{fig:outlier_methods}
\end{figure}

Our error detector (using relaxation events as an example) is trained on all samples in the ground-state ('0') cluster, which contains both ``true'' ground-state traces and relaxation errors. To construct this cluster, we first compute the \ac{MTV}~\cite{Benjamin2022} for all traces, defined as \( \mathrm{MTV}(Tr) = \frac{1}{|Tr|}\sum_{t=0}^{|Tr|-1} Tr(t) \), and then apply the \ac{KNN} method to identify relaxation errors. Since our ultimate goal is to develop an \ac{NN}-based detector that can be integrated into the main model, we use the \ac{KNN} outputs as pseudo-labels and train the detector to replicate these assignments.

However, we observe a significant overlap between the error traces filtered by the \ac{KNN} detector and those identified by the main model (Figure~\ref{fig:outlier_methods}). To avoid redundant computation, we instead feed the output of the main model into the detector. This allows the detector to focus solely on the hard-to-classify traces that the main model mislabels, rather than reprocessing the easy cases already handled correctly. As a result, the detector is able to achieve a better performance. Due to fabrication-related issues, qubit 2 suffers from excessive noise and crosstalk, so we exclude it from the detector pipeline. Using the \ac{KNN} method, we identify 2.4\%, 7.5\%, 3.4\%, and 2.9\% of traces from qubits 1, 3, 4, and 5 as relaxation events, and 0.6\%, 3.5\%, 2.4\%, and 0.7\% as excitation events. Since relaxation events dominate the misclassified traces, our detector mainly targets these cases.
\subsection{Main Classification Model}
\subsubsection{Challenge and Motivation}
As mentioned in Section~\ref{sec:intro}, existing \ac{ML}-based designs~\cite{Benjamin2022,Guo25,QubiCML,Liu25QR,mude2025efficient,MF/RMF,gautam2024low} are primarily based on \acp{FNN}. A key limitation of such architectures lies in their \textit{multiplicative scaling of parameters with the number of neurons per layer}. For qubit readout applications, achieving acceptable readout results requires readout durations between 600 and 1000 $ns$, which results in 600 to 1000 input features if no pre-processing is done. This means that for a readout duration of 1 $\mu s$, we can only reduce the model size to an absolute minimum of 1001 parameters if we directly map the input features to a single output neuron for binary classification. As soon as we introduce small hidden layers, the number of trainable parameters grows rapidly. Although recent approaches~\cite{Guo25,mude2025efficient} attempt to address the scaling issue, the intrinsic limitation of \acp{FNN} remains unchanged.
This multiplicative scaling issue, together with hardware implementation considerations, imposes strict limitations on their applicability to larger multi-qubit systems. \textbf{Therefore, it is desirable to develop a model whose size does not scale with the trace length.}
\subsubsection{Approach} 
We introduce the Mamba-based model for trace classification (as well as the error detector) as a replacement of \acp{FNN}, where the trace length does not directly influence the model size. Furthermore, Mamba introduces a linear-attention mechanism that allows the model to focus on or ignore specific time steps, which provides a similar functionality as the weight vectors produced by the \acp{MF}. Therefore, by using a Mamba-based model, we can reduce the model size significantly while maintaining a high readout accuracy.

\begin{figure}[t]
    \centering
    \includegraphics[width=\linewidth, trim=8mm 6mm 6mm 5mm, clip]{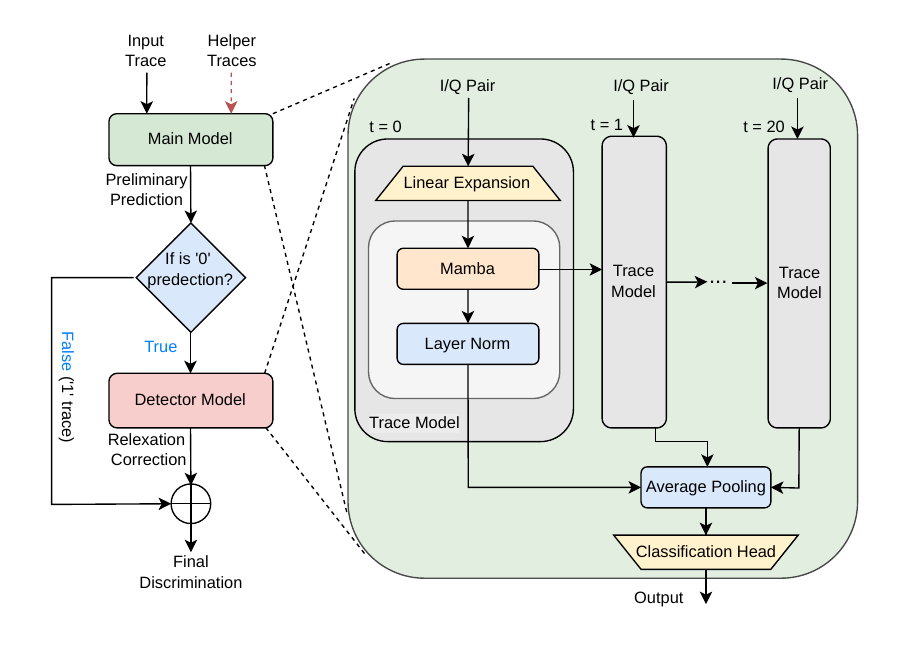}
    \caption{Architecture of the proposed multi-stage discriminator. The multi-stage workflow is shown on the left, while the right part illustrates the detailed Mamba architecture used in both the main classifier and the error detector.}
    \label{fig:workflow}
\end{figure}

Our proposed Mamba-based model, shown in the right part of Figure~\ref{fig:workflow}, consists of three parts at each time step. First, a linear layer expands the input dimensionality per time step from a single \ac{I} and a single \ac{Q} component to eight. While the Mamba block itself also has a parameter that controls the expansion of the input, our experiments show that we also need to add linear expansion at the input stage to increase performance. Since the input and output dimensionality of a Mamba block are equal, this simultaneously increases the output dimensionality. In the second part, the core Mamba block uses data from each timestep to evolve its internal hidden state and produce an output. To stabilize training, especially when using more than one Mamba block, our architecture includes a residual connection that gives us a gradient flow that bypasses the Mamba block. In the final part, we perform weighted pooling over all time steps and pass this vector of size eight through a linear classification head.

The main classification model is trained using Binary Cross Entropy Loss, while the detector model is trained using Focal Loss \cite{lin2018focallossdenseobject}, which emphasizes hard-to-classify samples. The values $\alpha$ and $\gamma$ allow the model to increase the weight of the minority class and focus on these hard-to-classify traces.
\begin{equation}
    FL(p_t) = -\alpha_t (1 - p_t)^\gamma \log(p_t)
\end{equation}
For the detector models, we employed a search algorithm to find the best $\alpha$ value. 

\subsection{Crosstalk Mitigation}

Excitation and relaxation errors are often caused by crosstalk between neighboring qubits during readout. To address this, we introduce optional helper traces to our main model to leverage these crosstalk effects for crosstalk mitigation. Previous research~\cite{Benjamin2022} has demonstrated that multi-qubit readout models can increase readout fidelity compared to single-qubit approaches. However, current multi-qubit readout approaches lack the flexibility to perform \ac{MCM}, as they typically rely on the simultaneous presence of all qubit traces. Due to the compact size of our Mamba-based network, we train two variants of the main model --- one without and one with helper traces --- both retaining scalability advantages over prior works. For the main model utilizing helper traces, we restrict the input to only the minimally required helper traces, avoiding the overhead of a full multi-qubit readout model. This strategy maintains a small model footprint and permits the use of the enhanced main model during \ac{MCM}, provided the necessary subset of traces is available.

\begin{figure}[tb]
    \centering
    \includegraphics[width=\linewidth]{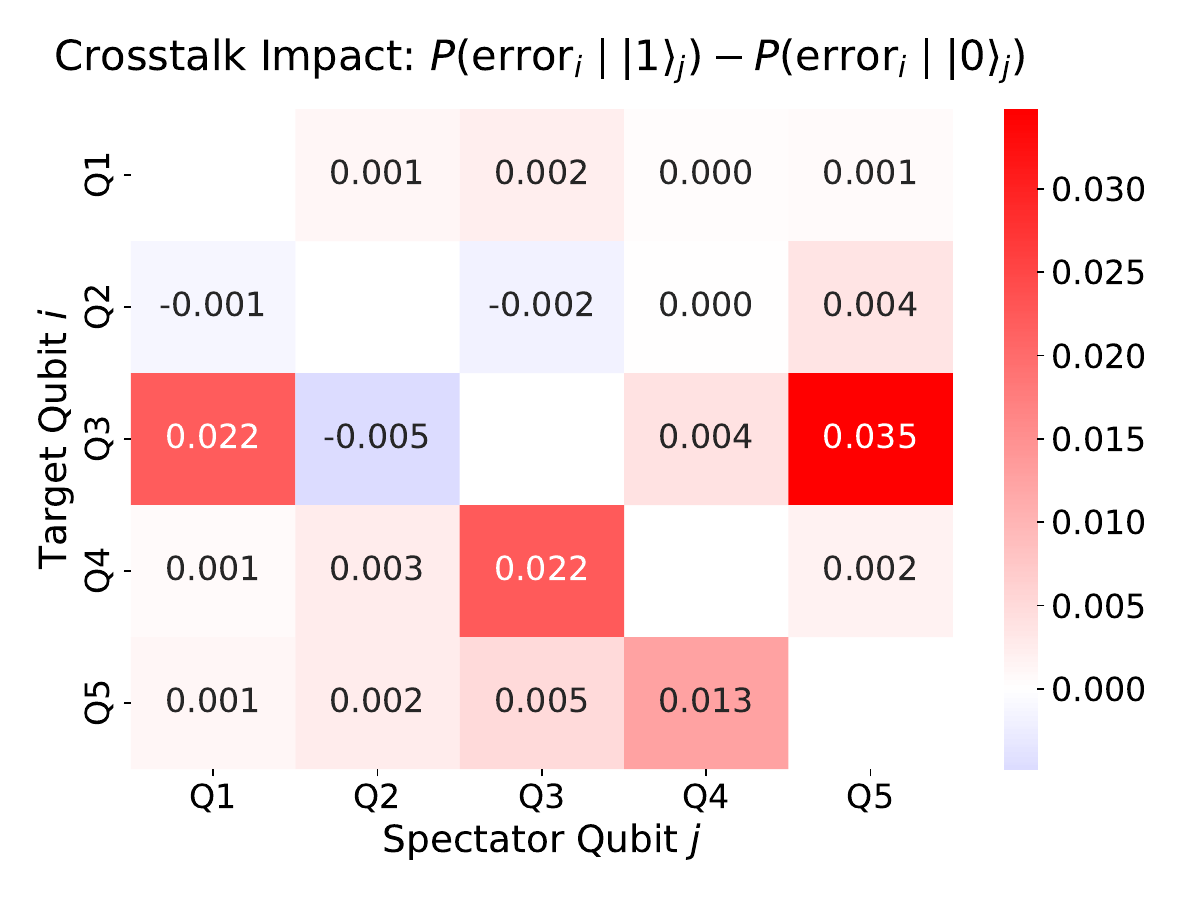}
    \caption{The heatmap illustrates the difference in error probability on a target qubit $i$ conditioned on the prepared state of a spectator qubit $j$. Positive values (red) indicate a higher error rate on qubit $i$ when qubit $j$ is prepared in the $|1\rangle$ state compared to the $|0\rangle$ state.}
    \label{fig:crosstalk}
\end{figure}

To identify which helper traces are strictly necessary for each qubit, we analyze the error probability of a target qubit conditioned on the preparation states of all other spectator qubits. Specifically, we evaluated the conditional error probabilities $P(\text{error}_i \mid |1\rangle_j)$ and $P(\text{error}_i \mid |0\rangle_j)$ of a target qubit $i$ when a spectator qubit $j$ is initialized in state $|0\rangle$ or $|1\rangle$. We then compute the difference $P(\text{error}_i \mid |1\rangle_j) - P(\text{error}_i \mid |0\rangle_j)$ to eliminate the base error probability of qubit $i$. These values are visualized in the matrix shown in Figure~\ref{fig:crosstalk}. The results indicate that certain spectator qubits increase the error probability of specific target qubits when prepared in state $|1\rangle$ instead of $|0\rangle$, highlighting the targeted crosstalk occurring between qubit pairs. For instance, the error rate of qubit 3 increases if qubit 1 or qubit 5 is prepared in state $|1\rangle$. Similarly, qubit 4 experiences higher errors if qubit 3 is in state $|1\rangle$, and qubit 5 is adversely affected when qubit 4 is in state $|1\rangle$. Qubits 1 and 2 show no significant dependencies under this metric.

Consequently, we train three additional main models: one for qubit 3 utilizing helper traces from qubits 1 and 5, one for qubit 4 utilizing the trace from qubit 3 as a helper trace, and one for qubit 5 utilizing the trace from qubit 4 as a helper trace. With these improved main models, we achieve relative improvements of 21.6\%, 7.1\%, and 3.2\% for qubits 3, 4, and 5, respectively, based on the Equation~\ref{eq:relative_improvement} where $F_{base}$ refers to the fidelity without helper traces. These results confirm that incorporating targeted helper traces effectively mitigates crosstalk-induced errors.

\begin{equation}
\text{Relative Improvement} = 
\frac{F_{\text{improved}} - F_{\text{base}}}{100\% - F_{\text{base}}}
\label{eq:relative_improvement}
\end{equation}

\subsection{End-to-end Workflow}

To produce the final classification results, we combine the main classification model (referred to as our lightweight model) with our error detector, both of which adopt the same Mamba architecture and follow the multi-stage workflow illustrated in Figure~\ref{fig:workflow}. We refer to this comprehensive workflow as our optimal model.

In the first stage, the main classification model provides an initial prediction, determining whether each trace corresponds to the ground state ('0') or the excited state (‘1’). If the main model predicts a ground state, the detector then tries to refine this prediction by classifying the sample as either a real ground-state trace or a relaxation event in the subsequent stage. When the detector identifies a relaxation trace for a sample initially predicted as ‘0’, the original prediction is corrected and reclassified as ‘1’. Notably, this ordering of the workflow is primarily motivated by the fact that the number of relaxation events is considerably smaller than that of normal traces. By structuring the process in this way, the error detector only needs to distinguish between '0' and relaxation errors, rather than among ‘0’, ‘1’, and relaxation events, which leads to more stable and accurate training.  


\begin{table}[tbp]
    \caption{Relative improvement in readout accuracy on qubits 1, 3, 4, and 5 achieved by the detector when paired with a lightweight FNN main model and our most space-efficient Mamba main model.}
    \centering
    \begin{tabular}{l|c|c|c|c}
        \hline
        \textbf{Model} & \textbf{Qubit 1} & \textbf{Qubit 3} & \textbf{Qubit 4} & \textbf{Qubit 5} \\
        \hline
        FNN & 1.6\% & 0.6\% & 0.3\% & 0.9\% \\
        \hline
        Mamba & 0.2\% & 0.1\% & 0.1\% & 0.1\%\\
        \hline
    \end{tabular}

    \label{tab:detector_results}
\end{table}

Depending on the performance of the main prediction model, this detector can achieve a relative improvement of the qubit's readout accuracy of up to 1.6\%, based on Equation~\ref{eq:relative_improvement}, where $F_{\text{improved}}$ and $F_{\text{base}}$ denote the readout fidelity with and without the detector, respectively. The detailed results are shown in Table \ref{tab:detector_results}. 

The results further indicate that the detector provides the greatest benefit when the main classification model struggles to reliably distinguish certain relaxation traces on its own. As the main model improves, the detector’s relative contribution naturally decreases. Nevertheless, even with our strongest main model, the detector still successfully corrects between 200 and 400 samples per qubit. However, due to the complexity of the detector's task, it also changes the predictions of some correctly classified traces, thereby reducing the net number of corrected traces. Therefore, as long as our detector predicts more true positives than false positives, it achieves an increase in readout accuracy.  

\begin{figure}
    \centering
    \includegraphics[width=1\linewidth]{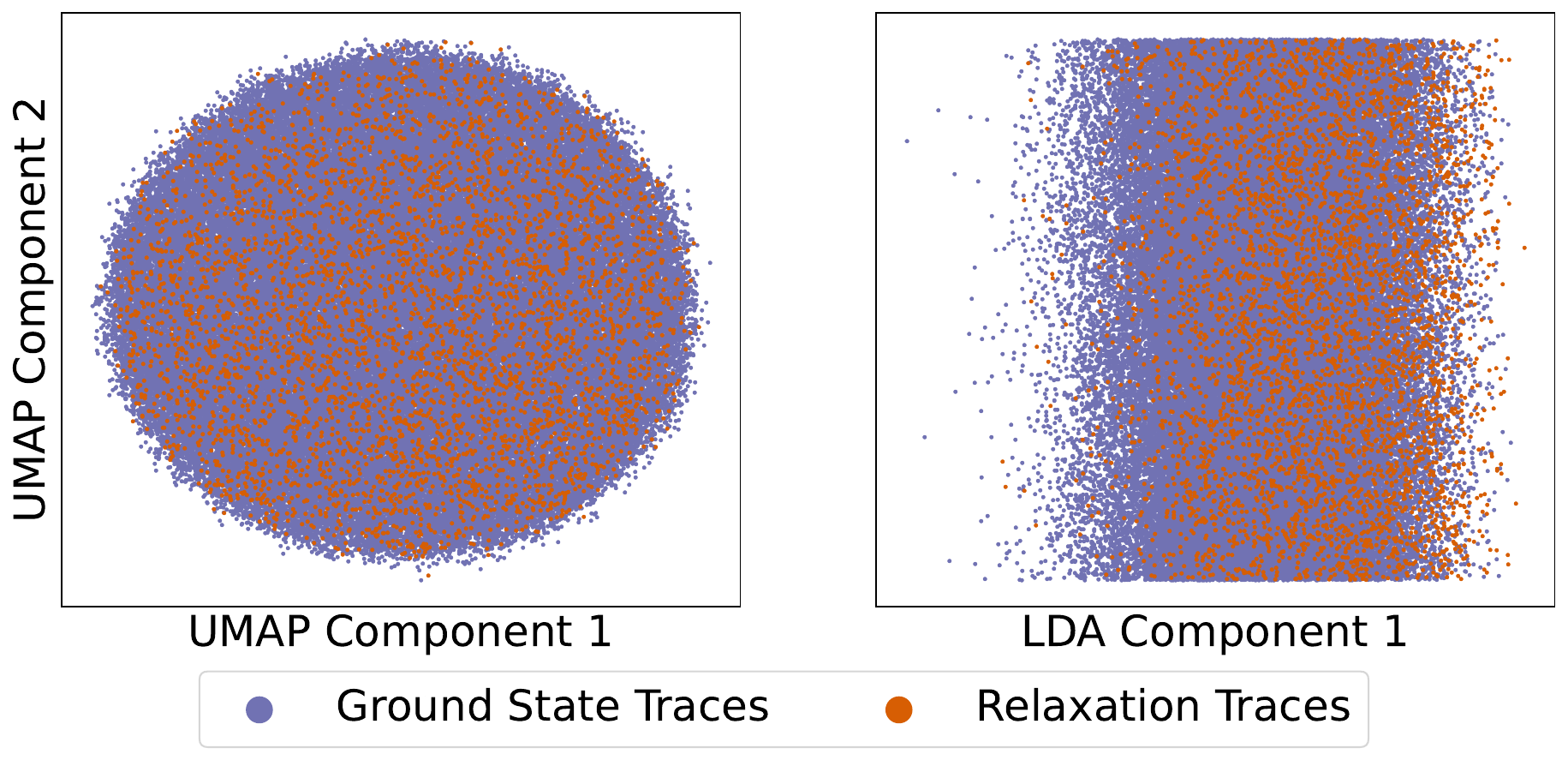}
    \caption{The UMAP clustering (left side) and the LDA results with a random offset on the y-axis (right side) show that there is no clear separation of true '0' traces and relaxation traces}
    \label{fig:main_model_predictions}
\end{figure}

The diminishing returns of the detector become clear from the UMAP and LDA visualizations in Figure~\ref{fig:main_model_predictions}, which shows that the traces classified as the ground state by the main model already form a compact cluster, with no distinct separation between misclassified relaxation traces and true ground-state traces. This is due to the fact that early relaxation events are nearly indistinguishable from true ground-state traces, making them especially difficult to correct to ‘1’. Such a phenomenon makes further performance improvements extremely challenging. 



\section{Evaluation and Experiments}
In this section, we present the main results of our multi-stage readout discriminator. We begin by introducing the quantum dataset and hardware used in our experiments. Then we focus on two key metrics for evaluating performance: model size (in terms of learnable parameters) and readout fidelity.
\subsection{Experimental Setup}
\subsubsection{Dataset}
We evaluate our work using the experimentally extracted \ac{QPU} readout traces comprising five qubits with frequency-multiplexed chip architecture~\cite{Benjamin2022}. These readout traces are extracted from the shared feedline and sampled by the \acp{ADC} at 500 MHz. The full readout trace is 2 $\mu s$, and each time point contains one \ac{I} and one \ac{Q} component. Due to the saturation behavior already observed in prior works, we evaluate our design using only the first 1~$\mu s$ of the readout trace.
For our single-qubit readout experiments, we demodulate the frequency-multiplexed readout trace to extract the readout signals for individual qubits. To further reduce noise, each 1$\mu s$ readout trace is subdivided into 50 $ns$ time bins, and the average \ac{I} and \ac{Q} values within each bin are computed, resulting in a trace consisting of 20 I/Q value pairs. 


Additionally, our dataset consists of 1.6 million multiplexed readout traces in total, containing 50,000 samples for each of the $2^5$ possible states for the five-qubit chip. For an individual qubit, this means that there are 800,000 samples for the ground state as well as 800,000 samples for the excited state. We then use a 70-15-15 split for our train, validation, and test set, resulting in 1.12 million traces for training and 240,000 traces each for validation and testing.
\subsubsection{Hardware}
The Mamba model is trained and evaluated on a Linux workstation equipped with NVIDIA H100 GPUs~\cite{h100} (100 GB memory each). 
\subsubsection{Discriminator Architecture}
To comprehensively evaluate the proposed architecture, we instantiate two model configurations representing different points on the accuracy–efficiency tradeoff. The first, denoted \textbf{Ours-Optimal}, incorporates all subcomponents introduced in Section~\ref{sec:Architecture}, targeting maximum readout fidelity. The second, denoted \textbf{Ours-Lightweight}, consists solely of the main Mamba model without helper traces, prioritizing parameter efficiency and scalability.

\subsection{Model Size}
Model size, in terms of learnable parameters, is an important metric to consider when evaluating scalability for large systems and applicability to the \ac{FTQC} era.
As introduced, the parameter count of the Mamba architecture no longer scales with the trace length; therefore, we can significantly reduce the model size. 

\begin{figure}
    \centering
    \includegraphics[width=1\linewidth]{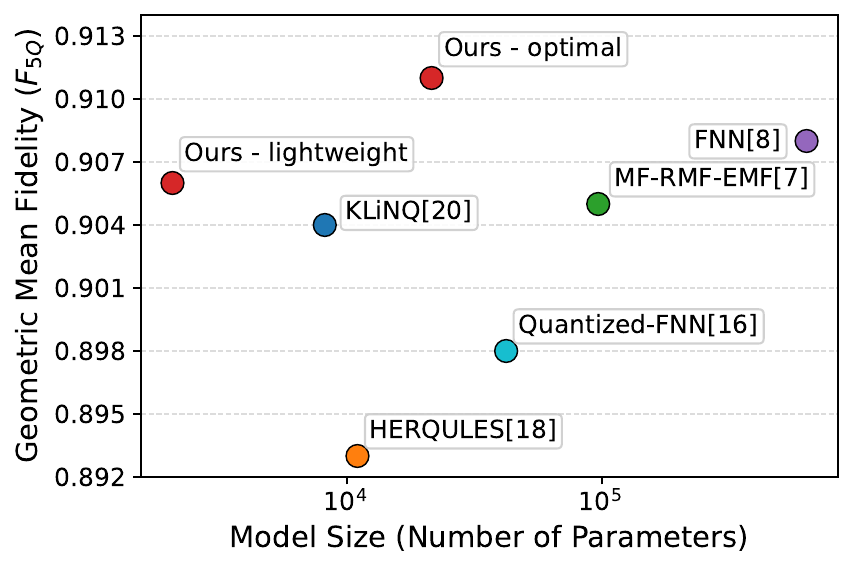}
    \caption{Comparison between model size and geometric-mean readout fidelity. The model size includes all trainable parameters, including parameters of \acp{MF} for HERQULES and MF-RMF-EMF.}
    \label{fig:size_vs_accuracy}
\end{figure}

Figure~\ref{fig:size_vs_accuracy} compares the model size and geometric mean fidelity, defined as $F_{5Q} = \sqrt[5]{F_1 F_2 \cdots F_5}$, for several related works. “Ours-Lightweight” contains only 769 parameters per qubit, making it the most lightweight design among all related works. Compared to the best-performing prior work, MF-RMF-EMF~\cite{mude2025efficient}, our approach not only achieves better discrimination performance but also reduces the model size by 49.6\%. Compared to KLiNQ~\cite{Guo25}, “Ours-Lightweight” requires 11.7\% more parameters for qubits 1, 4, and 5, but 78.5\% less for qubits 2 and 3. 

By employing a more powerful detector, our optimal model increases its parameter count to approximately 22,000 in total, slightly exceeding the sizes of HERQULES and KLiNQ. However, this configuration significantly enhances readout fidelity to the best among related works, while still using fewer parameters than the quantized FNN, MF-RMF-EMF, and the baseline \ac{FNN} (achieving a 98.0\% reduction compared to the latter). Additionally, excluding the detector stage while retaining the helper traces further reduces the total parameter count across all five qubits to roughly 11,000.


\subsection{Scalability Analysis}
As we scale superconducting quantum computers to the thousands of physical qubits required for \ac{FTQC}, the scaling behavior of the chosen readout architecture becomes critical. Control hardware, such as \acp{FPGA} and \acp{ASIC}, where this discrimination should be ultimately implemented, is resource-constrained and cannot scale arbitrarily. To achieve high readout fidelity in a frequency-multiplexed system, previous researchers leveraged inter-qubit correlations by proposing \ac{FNN}-based multi-qubit readout architectures. However, as mentioned, this leads to a quadratic scaling in input dimensionality and, consequently, model size.

Our architecture overcomes this bottleneck by aligning with the physical constraints of the quantum hardware. Frequency-multiplexed qubit systems often suffer from localized crosstalk because multiple qubits share a single feedline. Due to the frequency spacing required to avoid resonator collisions, there is a strict physical limit on how many qubits can share a single line. Therefore, the crosstalk stemming from frequency multiplexing occurs almost exclusively among a bounded, constant number of qubits within that specific feedline group. Because our approach only feeds the model the traces of the target qubit and its specific localized neighbors with significant crosstalk, the required number of parameters per target qubit remains constant. As a result, we achieve a strictly linear scaling in total model size with respect to the total number of qubits on the chip.

\subsection{Readout Fidelity}
\label{sec:RF}
Table~\ref{table:readout_comparison} summarizes the readout performance of our 
multi-stage discriminator, measured by $F_{5Q}$. 
“Ours-Optimal” strictly outperforms all state-of-the-art methods across 
all evaluation metrics. “Ours-Lightweight” also surpasses all existing 
methods, with the exception of the large baseline 
FNN~\cite{Benjamin2022}, which achieves comparable $F_{5Q}$ in the independent 
readout scenario despite having $99.4\%$ more parameters.

Specifically, compared with the MF-RMF-EMF method~\cite{mude2025efficient}, “Ours-Lightweight” achieves a higher readout fidelity across all qubits, with the exception of qubits 1 and 2, where the performance is comparable. Compared to KLiNQ~\cite{Guo25}, our approach achieves better performance for qubits 1, 4, and 5 while performing slightly worse for qubits 2 and 3. For qubits 2 and 3, the \ac{MTV} clusters for the ground and excited state overlap, making it harder for our detector and main model to distinguish relaxation traces from true '0' traces, similar as shown in Figure~\ref{fig:main_model_predictions}. In this table, we exclude QubiCML~\cite{QubiCML} due to its different dataset, and~\cite{gautam2024low} because it reports only one metric: an $F_{5Q}$ of 0.891.

\begin{table}[htbp]
\caption{Readout Fidelity Comparison For Single-Qubit (Independent) Readout Scenario.}
\label{table:readout_comparison}
\centering
\resizebox{\linewidth}{!}{%
\begin{tabular}{l|c|c|c|c|c|c|c|c}
\hline
\hline
\textbf{Method} & \textbf{Qubit 1} & \textbf{Qubit 2} & \textbf{Qubit 3} & \textbf{Qubit 4} & \textbf{Qubit 5} & \textbf{F$_{5Q}$} & \textbf{F$_{4Q}$} & \makecell{\textbf{Mid-}\\\textbf{Circuit}}\\ \hline
\hline
Baseline FNN & 0.969 & 0.748 & 0.927 & 0.947 & 0.970 & 0.908 & 0.953 & \ding{55}\\ \hline
HERQULES &  0.965& 0.730 & 0.908 & 0.934 & 0.953 & 0.893 & 0.940 & \ding{55}\\ \hline
KLiNQ & 0.968 & 0.748 & 0.929 & 0.934 & 0.959 & 0.904 & 0.947 & \ding{51}\\ \hline
MF-RMF-EMF & 0.971 & 0.745 & 0.923 & 0.939 & 0.969 & 0.905 & 0.950 & \ding{51}\\ \hline
\textbf{Ours-Lightweight} & 0.969  & 0.742 & 0.926 & 0.944 & 0.970 & 0.906 & 0.952 & \ding{51}\\ \hline
\textbf{Ours-Optimal} & 0.970  & 0.746 & 0.942 & 0.948 & 0.970 & 0.911 & 0.957 & \ding{51}\\ \hline
\end{tabular}
}
\end{table}


"Ours-Optimal", which incorporates helper traces, detector models, and a larger 
model capacity, achieves $F_{5Q} = 0.911$, outperforming all other models, 
including the much larger \ac{FNN}. This improvement is primarily attributed to 
the helper traces. As shown in Fig.~\ref{fig:crosstalk}, qubits 3 and 4 are most 
affected by crosstalk; by utilizing helper traces, the model can leverage 
additional context from neighboring qubits to correct a significant number of 
affected traces, resulting in notably high per-qubit fidelities of 0.942 and 
0.948 for qubits 3 and 4, respectively.

Beyond the helper traces, the remaining improvement can be attributed to the detector models. However, since our Mamba-based model 
is already highly capable, the detector has a comparatively smaller impact than 
it does on the \ac{FNN}, yielding only a marginal improvement in $F_{5Q}$.

To estimate the upper bound of achievable readout accuracy, we train and evaluate 
our model on a purified dataset in which error traces are removed using the 
\ac{KNN}-based approach described earlier. The resulting per-qubit fidelities are 
0.995, 0.745, 0.974, 0.992, and 0.970, giving $F_{5Q} = 0.930$ and 
$F_{4Q} = 0.983$. The persistently low fidelity of qubit 2 is attributed to 
known fabrication issues, which fundamentally limit its achievable performance 
regardless of the discrimination method.



\begin{figure}[htbp]
    \centering
    \includegraphics[width=1\linewidth]{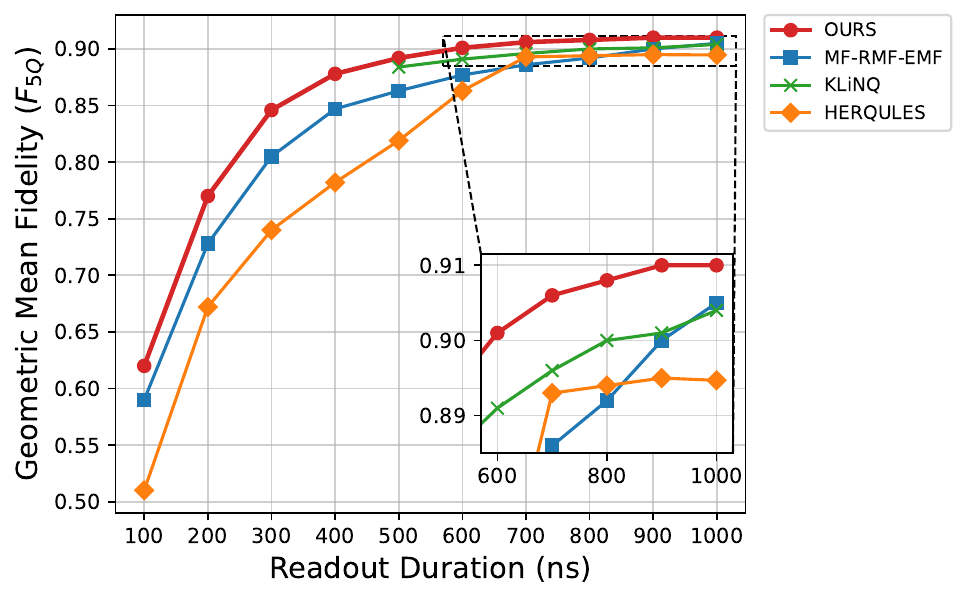}
    \caption{The cumulative readout accuracy (geometric mean) for different methods with different trace lengths. Since KLiNQ only provides the accuracy results up to 500 $ns$, our plot is truncated at this value. "OURS" refers to our optimal model.}
    \label{fig:cumulative_readout_accuracies}
\end{figure}

Furthermore, we evaluate how the cumulative accuracy of the best candidates from Figure~\ref{fig:size_vs_accuracy} varies with different trace lengths, as shown in Figure~\ref{fig:cumulative_readout_accuracies}. Notably, our model achieves a higher readout accuracy when trained on shorter readout traces compared to other solutions. For a readout duration of 600 $ns$, our multi-stage discriminator still achieves an acceptable fidelity of 0.896, which is similar to the accuracy obtained at 800 $ns$ by MF-RMF-EMF and HERQULES. For qubits 3 and 5, we already reach near-maximum readout accuracy of 0.926 and 0.970 at 700 $ns$ and 600 $ns$, respectively. For qubits 1, 2, and 4, we observe only a small loss in accuracy with 600 or 700 $ns$ readouts. Therefore, compared to the state-of-the-art works, we can further reduce the readout time by 100 to 200 $ns$, which is beneficial to \ac{MCM} for \ac{QEC}. Because the Baseline FNN requires over 600,000 trainable parameters per qubit, making it impractical and non-scalable. Therefore, we do not consider it as one of the best candidates.

\subsection{Case Study: Quantum Error Correction}

\begin{figure*}
    \includegraphics[width=0.95\linewidth]{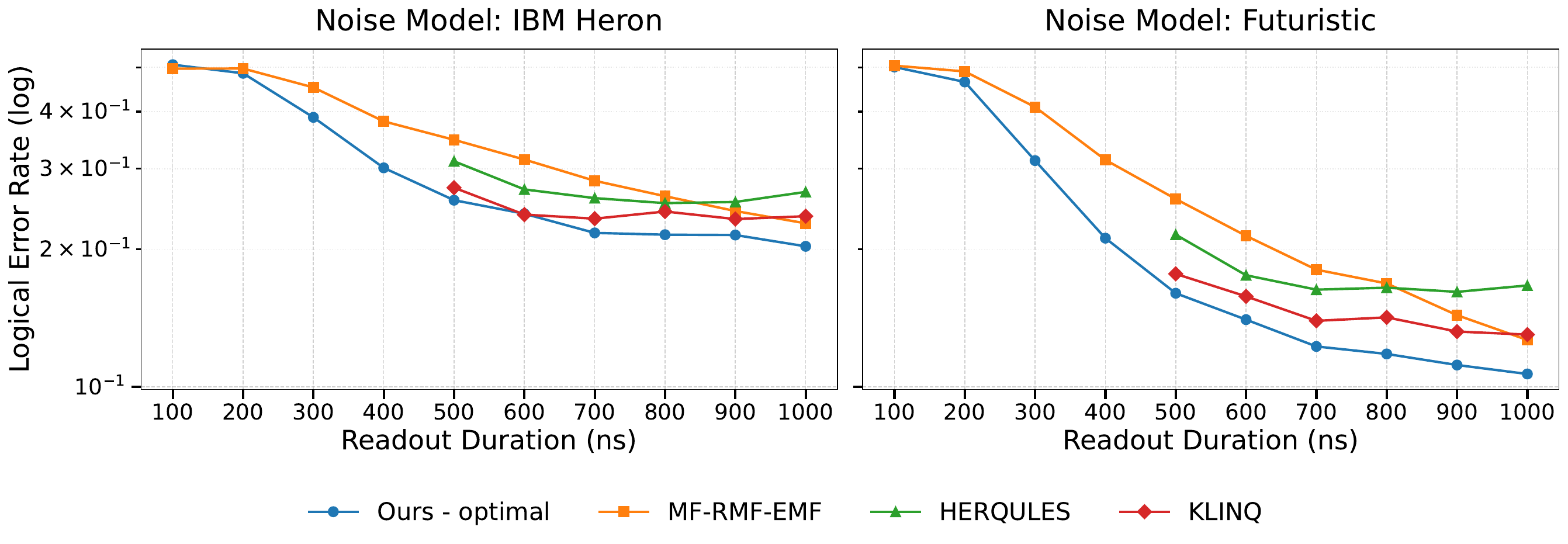}
    \caption{Simulated logical error rates of our proposed approach compared to MF-RMF-EMF, HERQULES, and KLiNQ (results available up to 500~$ns$ only) across varying readout durations. Results are shown after one round of quantum error correction using a rotated surface code ($d=7$) under two distinct noise profiles: an IBM Heron noise model (left) and a futuristic noise model (right).}
    \label{fig:logical-error}
\end{figure*}

While per-qubit readout fidelity provides a direct measure of discrimination accuracy, the ultimate figure of merit for \ac{FTQC} is the logical error rate. We therefore evaluate the downstream impact of our models on \ac{QEC} performance through Monte Carlo simulations. Since our experimental setup consists of five physical qubits, we simulate a rotated surface code with distance $d = 7$, grouping 97 physical qubits into a single logical qubit. We evaluate two distinct noise models: an IBM Heron noise model (assuming a two-qubit gate error of 0.5\%) and a futuristic noise model (assuming 10x lower error rates and an improvement in coherence time of 3$\times$). However, even though the readout fidelity would also improve under a futuristic noise model, we keep the readout fidelities at different readout durations fixed, as we can not accurately estimate the improvements in readout fidelity for all readout approaches. For each readout duration, one million shots are simulated per readout approach. The resulting logical error rates are plotted on a logarithmic scale in Figure~\ref{fig:logical-error}.



These simulations demonstrate that both our optimal and lightweight models consistently outperform all other evaluated solutions, including MF-RMF-EMF, HERQULES, and KLINQ, across both noise profiles. Our Mamba-based readout architecture achieves these lower logical error rates even at significantly reduced readout durations. As shown in Figure~\ref{fig:logical-error}, ``Ours-Optimal'' achieves strong performance at readout durations as short as 500 to 600 ns, effectively reducing the readout time by 100 to 200 ns compared to current methods. Specifically, in the current noise model (IBM Heron), our work reduces the logical error rate by 26\% at the readout duration of 500~$ns$, and the reduction is more pronounced (38\%) when we use the futuristic regime. This reduction in readout latency is crucial for QEC, as faster syndrome readout minimizes the time physical data qubits sit idle, reducing the time frame in which they can decohere. It enables faster QEC cycles and, therefore, improves QEC performance.


\subsection{Ablation Study}
We conduct an ablation study to evaluate the impact of multi-qubit information and input representation (demodulated vs. multiplexed traces) on readout performance.

Both prior works and our helper-trace experiments suggest that multi-qubit readout improves accuracy over single-qubit readout. Motivated by this, we also train two Mamba-based classification models (without the error detector), where one model uses multiplexed qubit readout traces and the other model uses five demodulated traces. By providing the model with information from all five qubits simultaneously, it can leverage inter-qubit correlations with all qubits to self-correct crosstalk–induced misclassifications, thereby achieving a higher readout fidelity of 0.748, 0.944, and 0.947 for qubits 2, 3, and 4, respectively. Contrary to our approach with helper traces, these models require the readout of all five qubits simultaneously.
However, when using the single-qubit readout architecture (i.e., with the same number of parameters) on the multiplexed traces, the performance drops drastically to an $F_{5Q}$ of only 0.756. Achieving a geometric mean fidelity of 0.906 with multiplexed traces requires a significantly larger model (extra 40,000 parameters), indicating that multi-qubit information without demodulation is only beneficial when the model has sufficient capacity to process the more complex multiplexed information.

In parallel, when we apply the single-qubit architecture to the five-demodulated-trace setting by expanding the input to five I/Q pairs, we obtain a slightly lower $F_{5Q}$ of 0.901. To achieve a similar performance of 0.906, the model size still needs to be increased, though not to the same extent as when using multiplexed raw traces. The readout fidelity for qubits 3 and 4 improves, showing the same results as our approach with helper traces. 

This observation provides a promising insight for future work on masked \ac{NN} architectures. Specifically, rather than training multiple distinct models for different helper-trace configurations, a single unified model could be employed. By masking the input traces of qubits that are not measured during a specific mid-circuit measurement, the network could be trained on random combinations of available traces. This approach would allow a single architecture to optimally leverage as much inter-qubit information as is available, maximizing readout fidelity without the overhead of maintaining multiple models.

\section{Related Work}
High-accuracy qubit-state discrimination remains an open problem, with a range of approaches proposed to maximize readout fidelity \cite{QubiCML, Guo25, mude2025efficient, MF/RMF, Benjamin2022}. Classical methods such as \ac{MF} and \ac{SVM} offer simplicity but limited accuracy, while more recent ML-based approaches,  including \ac{FNN} such as KLiNQ \cite{Guo25} and MF-RMF-EMF \cite{mude2025efficient}, achieve higher fidelity but suffer from parameter counts that scale multiplicatively with input size, limiting applicability to large multi-qubit systems.
While some of these models are effective for \textit{long} readout durations, they underperform for short ones, where fast \ac{MCM} feedback is most critical. To this end, we introduce a multi-stage discriminator based on the Mamba architecture, achieving linear complexity with respect to trace length and decoupling model size from trace length. Unlike prior multi-qubit methods \cite{Benjamin2022} that require simultaneous readout of all qubits and are therefore incompatible with MCM, our work supports single-qubit mid-circuit discrimination while optionally leveraging targeted helper traces to mitigate crosstalk in frequency-multiplexed architectures.

\section{Conclusion}
This work presents a multi-stage Mamba-based qubit state discriminator for superconducting quantum computers. The proposed architecture integrates a main 
Mamba classification model for initial state discrimination with an optional error detector that post-corrects relaxation errors, forming a seamless end-to-end pipeline. Using real sampled readout data, we demonstrate that our 
lightweight model achieves a geometric mean readout fidelity of 0.906 while 
reducing model size by 49.6\% to 99.4\% compared to state-of-the-art methods, with our optimal model further reaching 0.911. Both models maintain robust performance at readout durations as short as 500--600~$ns$, leading to up to a 26\% reduction in logical error rate for \ac{QEC} compared to the state-of-the-art. Furthermore, unlike existing multi-qubit readout methods, our single-qubit discrimination approach is fully compatible with \ac{MCM} and scales linearly with the number of qubits, making it well-suited for deployment on 
resource-constrained hardware platforms. Overall, this work demonstrates that lightweight, scalable, and low-latency qubit readout is achievable without sacrificing fidelity, representing a practical step toward \ac{FTQC}.

\bibliographystyle{IEEEtran}
\bibliography{ref}

\end{document}